\documentclass[aps,showpacs,twocolumn,superscriptaddress,nofootinbib,floatfix,section 10pt]{revtex4-2}
\usepackage{amsfonts,amssymb,stmaryrd,latexsym,amsmath,braket}
\usepackage{graphicx,subfigure}
\usepackage{comment}
\usepackage{times}
\usepackage{slashed}
\usepackage{bm}
\usepackage{braket}
\usepackage{appendix}
\usepackage{enumitem}
\usepackage{multirow}
\usepackage{array}

\usepackage[colorlinks=true,backref=true,linktocpage=true,citecolor=blue,urlcolor=blue,linkcolor=blue,pdfpagemode=UseOutlines]{hyperref}
\usepackage[dvipsnames]{xcolor}

\begin{document}
\title{Efficient vacuum state preparation for quantum simulation of strongly interacting local quantum field theories}
\author{Thomas D. Cohen}
\email{cohen@umd.edu}
\affiliation{Department of Physics and Maryland Center for Fundamental Physics, University of Maryland, College Park, MD 20742 USA}

\author{Hyunwoo Oh}
\email{hyunwooh@umd.edu}
\affiliation{Department of Physics and Maryland Center for Fundamental Physics, University of Maryland, College Park, MD 20742 USA}

\begin{abstract}
We present an efficient approach for preparing ground states in the context of strongly interacting local quantum field theories on quantum computers. The approach produces the vacuum state in a time proportional to the square-root of the volume, which is a square-root improvement in speed compared to traditional approaches. The approach exploits a novel method for traversing the path in parameter space in which the resources scale linearly with a path length suitably defined in parameter space. Errors due to practical limitations are controlled and do not exhibit secular growth along the path. The final accuracy can be arbitrarily improved with an additive cost, which is independent of the volume and grows slower than logarithmically with the overlap between the state produced and the exact ground state. We expect that the method could potentially hold practical value not only within the realm of quantum field theories but also in addressing other challenges involving long path lengths.
\end{abstract}

\date{\today}

\maketitle

Simulating aspect of quantum field theories (QFTs) on classical computers is often hindered by the notorious sign problem~\cite{Troyer:2004ge, PhysRevB.41.9301}; quantum computing is a candidate to evade the sign problem and simulate QFTs~\cite{Byrnes:2005qx, Jordan:2012xnu, Georgescu:2013oza, Zohar:2014qma, Bauer:2022hpo, Beck:2023xhh, Bauer:2023qgm, Funcke:2023jbq, DiMeglio:2023nsa}. Studying QFTs on quantum computers requires the preparation of initial states  
and various state preparation methods has been suggested~\cite{Ciavarella:2022qdx, Farrell:2023fgd, Roggero:2020sgd, Ciavarella:2021lel, Klco:2019xro, Moosavian:2019rxg, Buser:2020uzs, Stetcu:2022nhy, Choi:2020pdg, Lumia:2021tpu, HamedMoosavian:2017koz, Kane:2023jdo, Ball:2022dxy, Lee:2019zze, Davoudi:2022uzo, Brandao:2016mfe, deJong:2021wsd, Chi-Fang:2023edi, Holmes2022quantumalgorithms}. This paper focuses on the preparation of vacuum states of lattice field theories (with local interaction) aimed at future quantum computations. The class of theories considered are ones with finite correlation lengths and gapped spectra. The principal purpose of this paper is to show that the exact vacuum state of the (lattice) theories can be computed with arbitrarily high probability with a computational time that scales as the square root of the volume; an explicit algorithm doing this is developed.

In contrast to some problems in quantum information~\cite{Shor:1994jg, Grover1996AFQ}, in many physics applications it is essential to prepare a physical state a very large number of times: numerous properties of the state will be studied and the calculation of any of these require a large number of runs.
For these problems, it can be efficient to expend substantial resources the first time(s) the state is prepared in order to more efficiently prepare the state subsequently. The algorithm considered here is of this sort.

One general approach to state preparation is to begin with a system whose vacuum state can be explicitly constructed and then evolve the system (by varying a parameter, $\lambda$, specifying the Hamiltonian) in such a manner that---to a good approximation---the system remains in $|g(\lambda)\rangle$, the ground state of $\hat{H}(\lambda)$.  Adiabatic state preparation (ASP)~\cite{RevModPhys.90.015002, Wan:2020fwj, CoelloPerez:2021jkh, Ciavarella:2022skg, Chakraborty:2020uhf, Buser:2020uzs, Kovalsky:2022wcy, doi:10.1137/060648829} and methods based on the quantum Zeno effect (QZE) exploiting projection~\cite{boixo2010fast, 10.5555/2011804.2011811, PhysRevA.89.012314, PhysRevLett.108.080501, PhysRevA.63.052112} are in this category.  One can always choose parameterizations where $\lambda$ is dimensionless. 

There are two key issues in optimizing such algorithms.  An obvious one is the choice of path through parameter space, since clearly this affects how efficiently a ground state can be produced. The other is finding an efficient manner to traverse the path to find the ground state of interest. This paper addresses the second issue. 

There are important technical restrictions when using the approach outlined here:  for the algorithm to be efficient the theory must remains local for all values of $\lambda$ along the path and that no phase transitions (in the infinite volume theory) are encountered.  We recognize that the the restriction that no phase transitions are encountered along the path may reduce the applicability of this approach, but also note that this restriction applies generally to all methods based on traversing paths in parameter space including ASP and methods based on the QZE.

Even in the absence of phase transitions, there is a challenge faced by such methods for field theories with large volumes: the natural path through the various ground states is necessarily long---with path lengths scaling with the square root of the volume. It is important to note that while the size of the minimum spectral gap has received significant attention particularly in the context of adiabatic quantum computing~\cite{PhysRevA.81.032308, Wiebe_2012, 10.1063/1.2798382, Avron:1998th, farhi2000quantum}, the total cost of traversing a path in parameter space is also quite sensitive to the path length. This should be clear if one imagines a Hamiltonian trajectory that has been re-scaled so the the spectral gap is constant along the trajectory. Clearly, longer paths are more expensive to traverse. The approach developed here is designed to be efficient for long paths and should be useful for a variety of problems with long paths but it is particularly useful for QFTs with large volumes.

We follow a standard approach to define the path length. Denote
\begin{subequations}
$|g'(\lambda) \rangle \equiv \frac{ d|g(\lambda)\rangle }{d \lambda}$, where $|g(\lambda) \rangle$ is a differentiable function of  $\lambda$. Phases are chosen conventionally so that 
\begin{equation}
\forall \lambda_a, \lambda_b  \; ,  \; {\rm Im}\left [\langle g(\lambda_a)|g(\lambda_b \rangle \right ] =0 \;  . \label{Eq:conv}
\end{equation}
$L_{\lambda_a,\lambda_b}$---the path  length from $a$ to $b$  (where $b$ is a point further along the path than $a$) and total path length $L$---are then defined~\cite{10.5555/2011804.2011811, boixo2010fast, PhysRevA.89.012314} as
\begin{equation}
L_{\lambda_a,\lambda_b} \equiv \int_{\lambda_a}^{\lambda_b}  d \lambda \, \lVert | g'(\lambda) \rangle \rVert \; \;, \; \; L=L_{\lambda_0,\lambda_f} \;;   \label{Eq:L}
\end{equation}
\end{subequations}
$\lambda_0$ and $\lambda_f$ are initial and final values of $\lambda$ for the full path.
$L_{\lambda_a,\lambda_b}$ has a property of the path and is invariant under reparameterization of the path.

One needs to model the computational cost so that comparisons between different methods are straightforward. One important cost is the relative time needed by various schemes to complete the calculation on the same equipment;  this cost is denoted as $C$.  
$C$ depends on how the calculation is realized.  A useful and concrete way to envision it is to consider a hybrid between analog and digital quantum computers; the system undergoes continuous time evolution according to a Hamiltonian but quantum gates can control the circumstance for which this time evolution occurs and gated operations on  a state are allowable. This avoids issues associated with the cost to achieve the desired accuracy using Trotterization  while still allowing algorithms which require quantum control. Since rescaling the Hamiltonian by $A$ reduces the time by $1/A$, to make predictions of relative cost concrete, we assume that $H(\lambda)$ has always been scaled to the maximum that the equipment can accommodate.

Schemes directly based on the QZE\footnote{A straightforward calculation shows that for long paths, a QZE scheme with perfect projections and a negligible cost to (re)create the initial state yields an expected number of measurements, $N_p$, equal to $ N_p= L^2$. The energy-time uncertainty principle implies that the time to project the state of interest is of order $1/\Delta$, where $\Delta$ is the spectral gap; Eq.~(\ref{Eq:Lsq}) follows.} generically have 
\begin{equation}
    C \sim T \gtrsim \frac{L^2}{\Delta} \;, \label{Eq:Lsq}
\end{equation}  where $\Delta$ is the typical spectral gap along the path. Theorems for an upper bound on the time needed for ASP with fixed error~\cite{PhysRevA.81.032308, Wiebe_2012, 10.1063/1.2798382} are consistent with Eq.~(\ref{Eq:Lsq}). 

As will be shown, the path length is proportional to the square-root of the volume of the system:
\begin{equation}
  \frac{L_{V_2}}{L_{V_1}} = \sqrt{\frac{V_2}{V_1}} \; .  \label{Eq:LV}
\end{equation} 
The cost of schemes based on QZE or ASP then scale linearly with the volume. In this paper, a scheme with $C \sim \sqrt{V}$ is developed.

\begin{widetext}
Let us derive Eq.~(\ref{Eq:LV}). Locality of the interactions and the finiteness of the correlation lengths imply that at large volumes 
\begin{subequations}
\begin{equation}
\langle g'(\lambda_0)|g'(\lambda_0)\rangle = -\left .\frac{\partial^2 \langle g(\lambda_1 ) |  g(\lambda_0 )\rangle}{\partial (\lambda_1)^2} \right|_{\lambda_1=\lambda_0}\!\!\! =\left(\frac{V}{\xi^d} \right) h(\lambda_0) \times \left(1+{\cal O}\left (\frac{\xi^d}{V} \right) \right),  \label{Eq:OL1}
\end{equation}
where $h$ is independent of the volume at large $V$, $\xi$ is the correlation length, and $d$ is the dimension of space. More generally,
\begin{equation}
\left .\frac{\partial^n \langle g(\lambda_0  +\Delta \lambda/2 ) |  g(\lambda_0 - \Delta \lambda/2)\rangle}  {\partial (\Delta \lambda)^n}  \right|_{\Delta \lambda = 0} = - \frac{1+(-1)^n}{2}  \; 
 \frac{\langle g'(\lambda_0)|g'(\lambda_0)\rangle^\frac{n}{2}n!}{2^{n/2} \,(n/2)!} \;
\left (1 +{\cal O}\left (\frac{\xi^d}{V} \right) \right),  \label{Eq:OL2}
\end{equation}
implying
\begin{equation}
    \langle g(\lambda_1) |  g(\lambda_0) \rangle =  e^{ -L^2_{\lambda_0 , \lambda_1}/2} \left(1+{\cal O}\left (\frac{\xi^d}{V} \right) \right ) \; \; {\rm and} \; \;
    \lVert |g'(\lambda_0+ \Delta \lambda )\rangle \rVert = \lVert |g'(\lambda_0)\rangle \rVert \times  \left(1+{\cal O}\left (\frac{\xi^d}{V} \right)^{\frac 12} \right) \; , \label{Eq:OL3}
\end{equation} \label{Eq:OL}
\end{subequations}
\end{widetext}
which in turn implies that $L \sim \left(\frac{V}{\xi^d}\right )^{\frac 12}$. 

The scheme proposed here differs fundamentally from ASP (where the system passes through every ground state from the initial one, $|g(\lambda_0)\rangle$, to the final one, $|g(\lambda_f)\rangle $).  It also differs fundamentally from conventional QZE schemes where only discrete points along the path, specified by $\lambda_i$, are considered, but adjacent points on the path are close:  $|\langle g(\lambda_{i+1})| g(\lambda_i)\rangle|^2 \approx 1$.  

Like QZE-based schemes, our approach uses discrete points along the path.  However, instead of large overlaps between neighboring points on the path, ideally the points are chosen so that 
\begin{equation}
    |\langle g(\lambda_{i+1})| g_i(\lambda_i )\rangle|^2=1/2 \; , \;   L_{\lambda_{i},\lambda_{i+1}} = \sqrt{\log (2)} \; ;
\label{Eq:half}\end{equation}
the second form is valid for large volumes.  
We assume that in initial studies, the set of points satisfying Eq.~(\ref{Eq:half}) have been found (to a good approximation).  

\begin{figure*}[t]
    \includegraphics[width=1\textwidth]{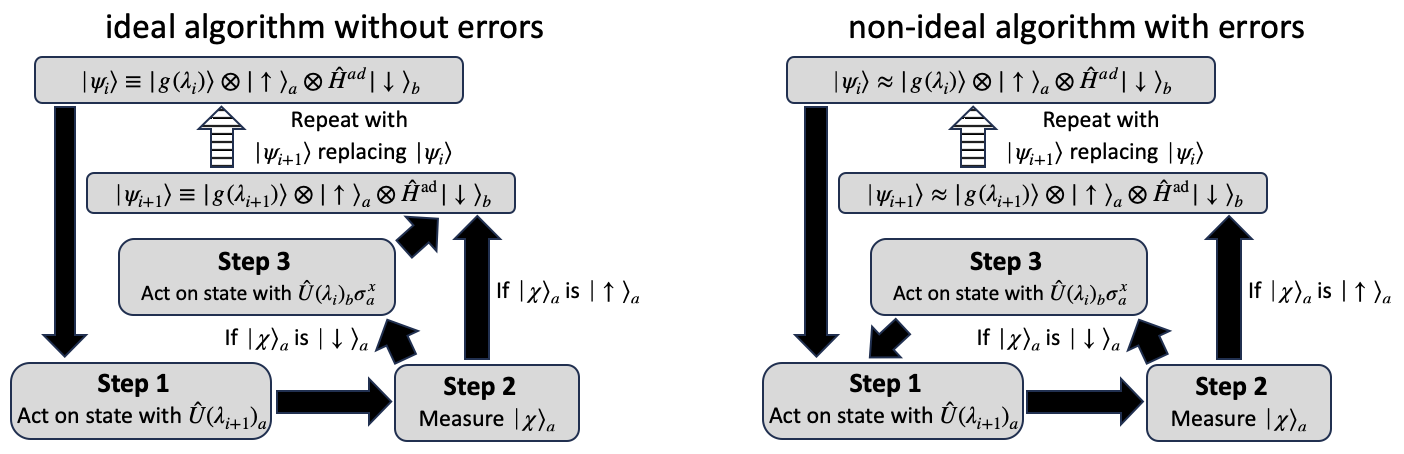}
    \caption{A flowchart for one cycle in the idealized  algorithm, which assumes no errors (left), and a realistic version that can handle errors (right). }
    \label{fig:flowchart}
\centering \end{figure*}

An idealized version of the algorithm invokes an oracle, $\hat{U}(\lambda)_\gamma$, parameterized by $\lambda$ and $\gamma$, where $\gamma$ can take the value of $a$ or $b$. 
$\hat{U}(\lambda)_\gamma$ acts on a state combining $|\psi \rangle$, the state describing the system of interest, with two ancillary qubits $|\chi \rangle_a$ and $|\chi \rangle_b$. The action of $\hat{U}(\lambda)_\gamma$ is given by:
\begin{equation}
 \begin{split}
 \hat{U}(\lambda)_a & = |g(\lambda)\rangle \langle g(\lambda)| \otimes \hat{I}_a \otimes \hat{I}_b \\ 
 & + \left (\hat{I}_\psi - |g(\lambda)\rangle \langle g(\lambda)|  \right ) \otimes \hat{\sigma}^x_a \otimes \hat{I}_b \; ,\\
  \hat{U}(\lambda)_b & = |g(\lambda)\rangle \langle g(\lambda)| \otimes \hat{I}_a \otimes \hat{I}_b \\ 
 & + \left ( \hat{I}_\psi - |g(\lambda)\rangle \langle g(\lambda)| \right ) \otimes \hat{I}_a \otimes \hat{\sigma}^x_b \; ,
\end{split} \end{equation}
where $\hat{\sigma}^x_a$ and  $\hat{\sigma}^x_b$ are Pauli operators acting on the ancillas,  $\hat{I}_a$  and  $\hat{I}_b$ are identity operators acting on the ancillas, while $\hat{I}_\psi$ is the identity acting on the state of the system.

$\hat{U}_a(\lambda)$ flips $| \chi \rangle_a$, if the system is in an excited state and leaves it if the system is in the ground state; $\hat{U}_b(\lambda)$ does the same for $| \chi \rangle_b$. Highly accurate approximations of the oracle can be constructed using, for example, Kitaev's phase estimation scheme~\cite{Kitaev:1995qy} or a variant of the rodeo algorithm or similar approaches~\cite{Choi:2020pdg, Cohen:2023rhd, Bee-Lindgren:2022nqb, Stetcu:2022nhy}. It is straightforward to show that when $|e \rangle$ is a superposition of excited states of $H(\lambda)$ and $|\alpha|^2 + |\beta|^2 =1$ 
\begin{equation}
\begin{split}
 \hat{U}(\lambda)_b & \left( \left ( \alpha |g(\lambda) \rangle + \beta |e \rangle\right) \otimes |\chi \rangle_a  \otimes \hat{H}^{\rm ad}|\downarrow \rangle_b \right ) = \\
& \left ( \alpha |g(\lambda) \rangle - \beta |e \rangle\right) \otimes |\chi \rangle_a  \otimes \hat{H}^{\rm ad}|\downarrow \rangle_b \; ,
\end{split} 
\end{equation}
where $\hat{H}^{\rm ad}$ is the Hadamard operator; the effect of $\hat{U}(\lambda)_b$ is to reflect any state in the physical Hilbert space around $ |g(\lambda) \rangle$.

Consider an idealized case: the oracle is implemented exactly and the points $\lambda_i$ are known exactly. Then a simple scheme acts to move a system in the ground state at point $\lambda_i$ to the ground state at point $\lambda_{i+1}$. The scheme requires that $| \chi \rangle_a$ begins as $|\uparrow\rangle_a$ while $|\chi \rangle_b$ begins as $\hat{H}^{\rm ad}|\downarrow \rangle_b$; the full state is $|\psi_i\rangle=|g(\lambda_i)\rangle \otimes |\uparrow\rangle_a \otimes \hat{H}^{\rm ad}|\downarrow\rangle_b \;$.

\begin{figure}[b]
 \includegraphics[width=.37\textwidth]{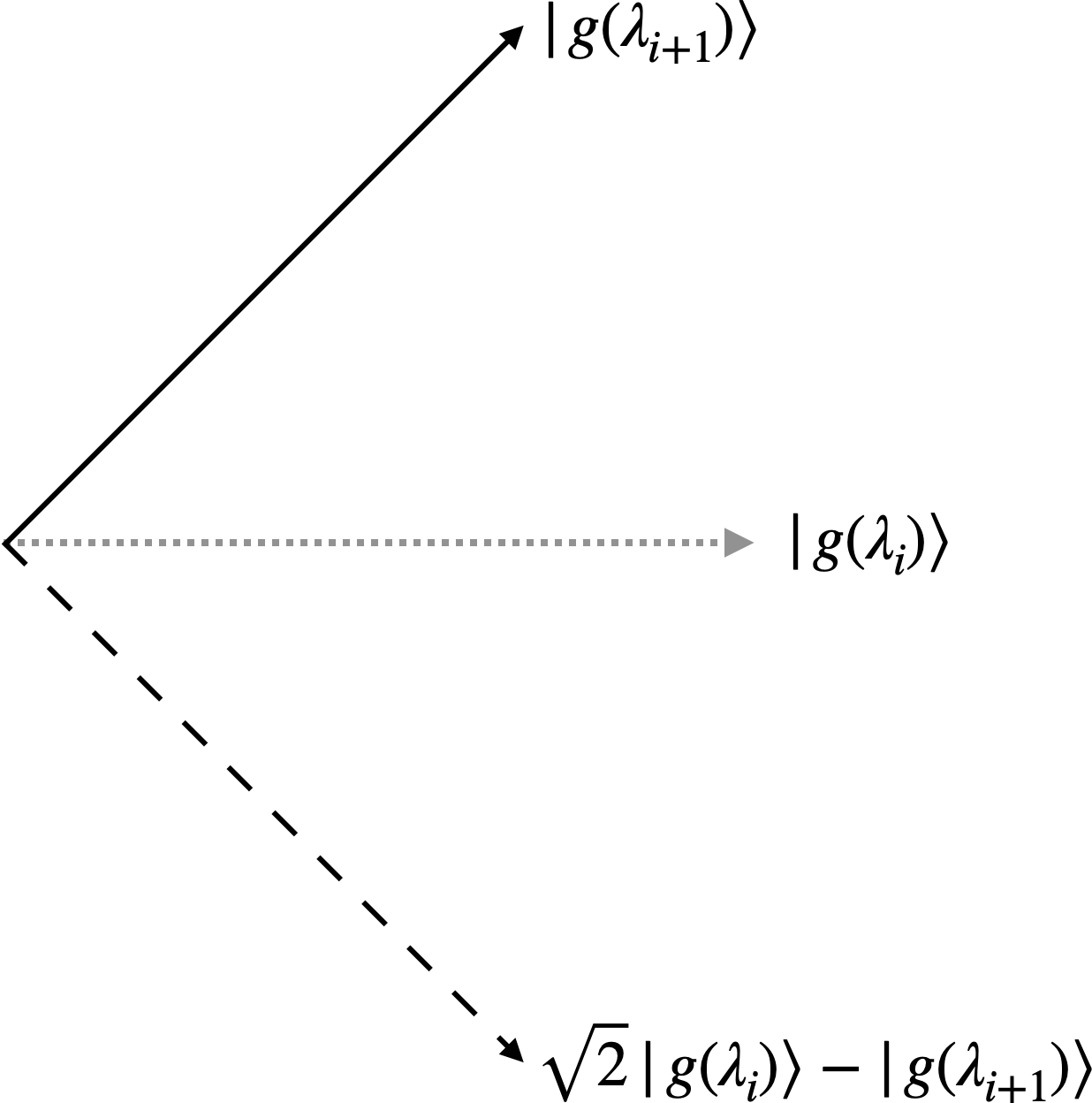}
    \caption{The two-dimensional space spanned by $|g(\lambda_i) \rangle$ and $|g(\lambda_{i+1}) \rangle$.}
    \label{fig:vectors}
\centering 
\end{figure}

A flowchart for the idealized algorithm is on the left in Fig.~\ref{fig:flowchart}. The first step is to act on $| \psi_i\rangle$ with $\hat{U}(\lambda_{i+1})_a$. Next, $| \chi \rangle_a$ is measured. Half of the measurements will yield $|\uparrow\rangle_a$, indicating that the system is in the ground state of $\hat{H}(\lambda_{i+1})$; nothing more needs to be done: the system has advanced to the next point along the path. If the measurement yields $|\downarrow\rangle_a$, the system is in an excited state of $\hat{H}(\lambda_{i+1})$;  this excited state is in the 2-dimensional vector space spanned by $|g(\lambda_{i})\rangle$ and $|g(\lambda_{i+1})\rangle $; the state is $\sqrt{2}|g(\lambda_{i}) \rangle - |g(\lambda_{i+1}) \rangle$, as seen Fig.~\ref{fig:vectors}. When $|\chi \rangle_a$ is measured to be $|\downarrow \rangle_a$, the third step is implemented: $\hat{U}(\lambda_{i})_b \sigma_a^x$ acts on the state. As seen in Fig.~\ref{fig:vectors}, this reflection yields $|g(\lambda_{i+1})\rangle$. Regardless of whether $|\chi\rangle_a$ is measured to be $|\uparrow\rangle_a$ or $|\downarrow\rangle_a$, after the cycle the system is in the ground state of $\hat{H}(\lambda_{i+1})$, having called the oracle at most two times. 
In this idealized scenario, a path of length $L$ would be traversed by calling the oracle $3 L/\sqrt{4\log(2)}$ times (up to small statistical correction) when $L \gg 1$. If each call of the oracle took a fixed time, then the total time would scale linearly with $L$, and the cost of state preparation scales as $\sqrt{V}$.

This ideal case is unrealistic: there will always be errors arising from imperfect information, imperfect devices or the inability to construct a perfect oracle that runs in finite time (even with perfect devices and perfect information). There have been previous schemes~\cite{10.5555/2011804.2011811, boixo2010fast, PhysRevA.89.012314} proposed for the traversal of long paths that mitigate effects of such imperfections by increasing the resources at each stage so that the errors could be kept below the final desired value at all points along the path; such schemes lead to additional costs and they scaled asymptotically with $L$ no better than $L \log L$. A key insight in this work is that it is sufficient that there is no secular growth in the error with increasing $L$; at the end of the path, all one needs is that admixtures of excited states that are either strictly bounded independent of L or given in terms of a sufficiently narrow distribution whose width does not grow with $L$. If so, one can then use an ``afterburner'' to produce the ground state with whatever accuracy is needed: vacuum states are calculable with a cost that scales as $V^{1/2}$ without a logarithmic correction. 

The flow chart on the right in Fig.~\ref{fig:flowchart} is a variant of the algorithm that is useful for non-ideal conditions and allows long paths to be traversed without a secular growth of errors. The scheme makes a single change from the ideal case: after step three, the algorithm returns to step one rather than completing the cycle. If everything were ideal, this change would be redundant: the system would  be in the ground state of $\lambda_{i+1}$ and steps one and two would simply go on to the next cycle. However, under non-ideal conditions it suppresses secular growth of errors with only a modest multiplicative increase in cost relative to ideal case.

There are several causes of errors. A key one is that one cannot project perfectly in finite time. Imperfect projections lead to ``false positives'' where an excited state component of a wave function is identified as the ground state. A second factor is the inability to prepare points that satisfy Eq.~(\ref{Eq:half}) exactly. Additionally, there can be ``false negatives'' in projection: uncertainty in knowledge of the ground state energy can lead to the projection indicating that the system is not in the ground state even when it is. 

One might anticipate that these errors could accumulate along the path, causing the amplitudes of ground states to diminish along the path. However, a simple heuristic argument suggests that the error does not grow in a secular manner as the path is traversed.

Denote the maximum error introduced in any stage of the algorithm as $\epsilon$; it represents the maximum amplitude introduced for non-ground state components at any stage of the algorithm (assuming it had started in the ground state). Given that each stage concludes with the oracle $\hat{U}(\lambda)_a$, the error introduced in the $n^{\rm th}$ iteration will be mitigated by the $(n+1)^{\rm th}$ iteration. Of course, the $(n+1)^{\rm th}$ iteration will also introduce an error (which will be less than or equal to $\epsilon$). The same situation happens for next iterations and consequently, the error converges with a geometric series: the error will not grow secularly.
 
To test this heuristic reasoning, we implemented the algorithm numerically in a ``toy model''. In constructing this model, we exploited the fact that what matters is the ground state itself (along with excited states) rather than the Hamiltonian that leads to the states. Therefore, for simplicity, the model consists of ``ground'' and ``excited'' states constructed (essentially randomly) directly as linear combinations of an orthogonal basis of 1024 levels (this number was chosen to be large enough to represent a generic system but small enough to keep the calculation tractable). The inner product between the $i^{\rm th}$ and $(i+1)^{\rm th}$ ground state was chosen to be close to $1/\sqrt{2}$ for each step. To simulate false positive errors we assumed that the projection allowed through excited states with a probability of 0.1 or less (randomly chosen with a uniform distribution); this is extremely conservative (a single ``super iteration'' of Rodeo projection has maximum rate of less than 0.05 and is typically much less~\cite{Cohen:2023rhd}). We also chose a 0.05 probability of a false negative (which is also quite conservative if one has reasonable knowledge of the ground state energies) and assumed that $\cos^{-1}$ of the inner product between $i^{\rm th}$ and $(i+1)^{\rm th}$ ground states was within $10 \% $ of $\frac{\pi}{4}$ (which again should be quite achievable).

\begin{figure}[t!]
\centering
 \includegraphics[width=0.48\textwidth]{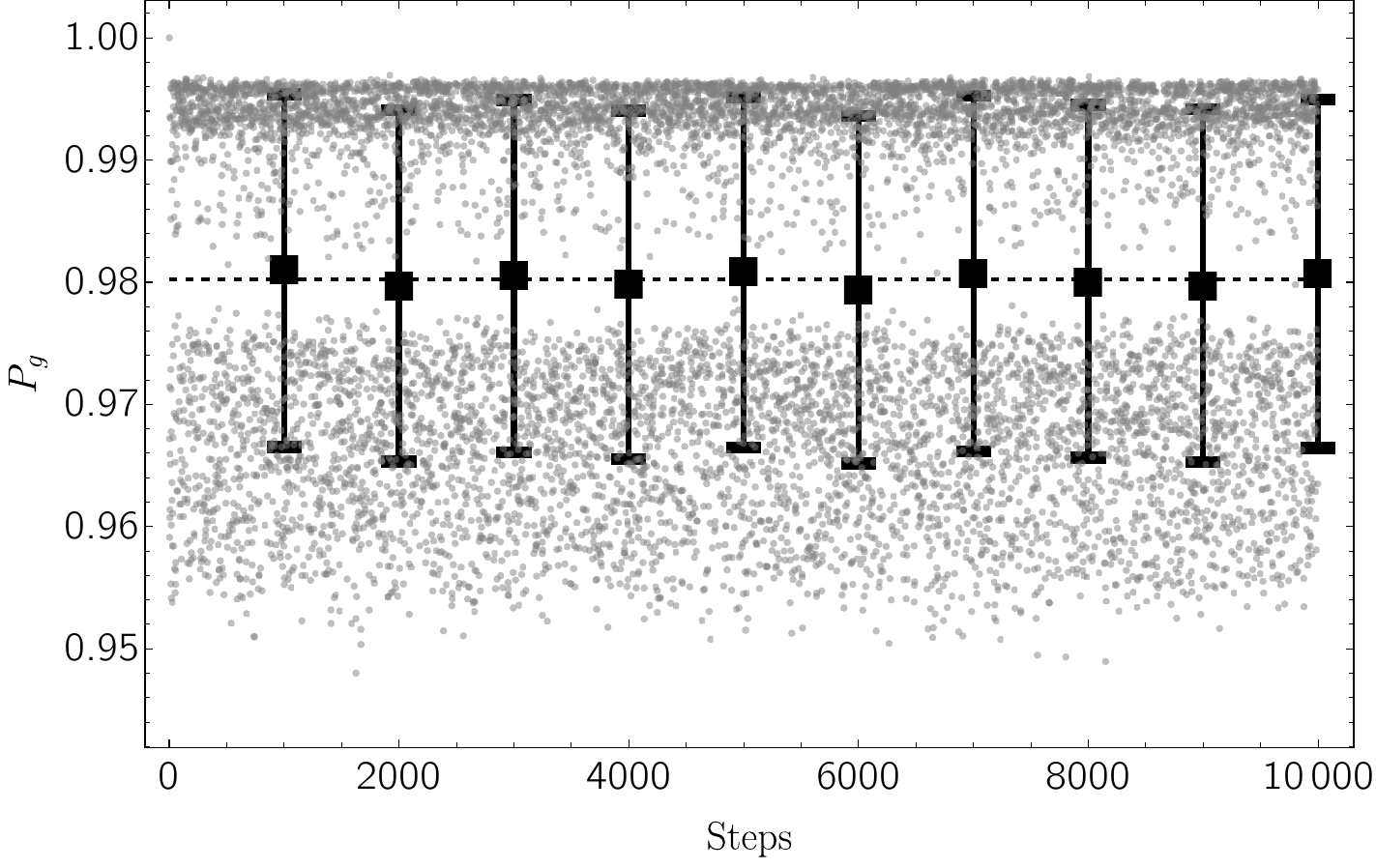}
    \caption{A numerical simulation of the algorithm given for the algorithm in Fig.~\ref{fig:flowchart} using the toy model discussed in the text. $P_g$ represents the ground state probability (the square of the inner product of the true intermediate ground state with the one produced by the algorithm). The dots represent the ground state probabilities at each step, while the rectangles represent averages taken over every 1000 steps (with the standard deviations indicated by the error bars). The dashed line is the average of all steps.  
    Note that the scale of $P_g$ is highly compressed, ranging from 0.94 to 1.00.}
    \label{fig:numerical}
\end{figure}

The results of a simulation using this toy model are shown in Fig.~\ref{fig:numerical}:  In that figure, $P_g$, the ground state probabilities (the square of the intermediate ground state component of the state produced at a given step along the path) are nearly unity for all steps (with a minimum $P_g$ of greater than 0.947). Although errors fluctuate as additional steps along the path are taken, the errors do not grow in a secular manner:  The characteristic distribution of $P_g$ for the steps 9000-10000 is qualitatively the same as steps 1-1000, with both the average and standard deviations nearly identical. This numerical check strongly suggests that the conclusion of the heuristic argument is sound.

There is strong evidence that the errors do not grow as the path is traversed. Assuming that this is true, the inner product of the state produced at the end of the long path, and the true ground state will be close to unity. If one wishes the final state to be extremely close to the true ground state, one can use an "afterburner". It involves the last step of the traversal---which is likely not very close to satisfying Eq.~(\ref{Eq:half})---and thus may require several rounds of projection and reflection but is guaranteed to converge. It simply uses a projector with a false positive rate which is exponentially smaller than one used to traverse the path. This ensures that when the last step converges one has the true ground state with exponential accuracy. As was shown in~\cite{Cohen:2023rhd}, the cost in time grows more slowly than logarithmically with the false positive rate when using the Rodeo algorithm. Thus, for long paths the total cost is dominated by traversing the path and not the afterburner.

The total cost for long paths depends on the cost of each projection as well as the number of projections needed. Since typical projection methods require a time proportional to the inverse of the spectral gap, $\Delta$, the total cost is given by $C \sim V^{\frac12} \left \langle \frac 1{\Delta} \right \rangle $
where $\langle \Delta^{-1}\rangle$ is the average of the inverse spectral gap along the path.

Properties of the path need to be determined in initial traversals of the path so that subsequent traversals can be efficient: reasonable knowledge of the values of $\Delta \lambda$ to reach the next point approximately satisfying Eq.~(\ref{Eq:half}), the ground state energy (to suppress the false negative rate) and the spectral gap (to efficiently project) are needed at each point on the path. The algorithm outlined above is applicable to all problems with long path lengths. However, if determining the properties of the system along the path during initial studies is too expensive, the algorithm may be impractical.   

Fortunately, for local field theories with large volumes relevant quantities change very slowly from $\lambda_i$ to $\lambda_{i+1}$; one can extract the values at relatively few points and then smoothly interpolate. To see this, recall that $\Delta \lambda_i=\lambda_{i+1}-\lambda_i \propto V^{-\frac12}$ from Eq.~(\ref{Eq:half}). Moreover, at large volume, the spectral gap, $\Delta(\lambda)$, depends only on $\lambda$ and not the volume. Therefore, $\Delta \lambda_{i+N}$ can be approximated as $\Delta \lambda_{i}$ up to $N \propto V^{\frac12}$. Similarly, Eqs.~(\ref{Eq:OL1})-(\ref{Eq:OL3}) implies that $\Delta(\lambda_{i+N})$ is well approximated as $\Delta(\lambda_i)$ until $N \propto V^{\frac12}$. However, since the ground state energy, $E_0(\lambda)$, is proportional to the volume, $E_0(\lambda)$ must be approximated as a Taylor series; an $n^{\rm th}$ order Taylor series at $\lambda_i$ provides an accurate estimate of $E_0(\lambda_{i+N})$ until $N \propto V^{\frac12-\frac1n}$.

In this work, a highly efficient state preparation method on quantum computers was presented, particularly tailored for use with QFTs with large volumes.
It was demonstrated that errors do not accumulate secularly. 
The results here are of interest from a theoretical quantum information perspective as they indicate that the cost of preparing the ground state of QFTs scales as the $V^{\frac12}$. The approach may also prove to be valuable practically both in the context of QFTs and perhaps in other problems with long path lengths.  However, practical implementations of the approach are unlikely in the short term; the approach relies on controlling the buildup of errors and is unlikely to be viable without the development of fault-tolerant quantum computers.

\begin{acknowledgments}

This work was supported in part by the U.S. Department of Energy, Office of Nuclear Physics under Award Number(s) DE-SC0021143, and DE-FG02-93ER40762.

\end{acknowledgments}

\bibliography{refs.bib}


\end{document}